\documentclass[onecolumn,prd,showpacs]{revtex4}
\usepackage{graphicx}
\usepackage{amssymb}
\begin{document}

\title{\bf Anti-$\mathcal{PT}$ Transformations And Complex Non-Hermitian $\mathcal{PT}$-Symmetric Superpartners}
\author{Taha Koohrokhi}
\email{t.koohrokhi@gu.ac.ir}
\affiliation{Department of Physics, Faculty of Sciences, Golestan University, Gorgan, Iran}

\author{Sehban Kartal}
\email{sehban@istanbul.edu.tr}
\affiliation{Istanbul University, Department of Physics, 34000, Istanbul, Turkey}

\author{Ali Mohammadi}

\begin{abstract}
We propose a new algebraic formalism for constructing complex non-Hermitian $\mathcal{PT}$-symmetric superpartners by extending a conventional shape-invariant superpotential into the complex domain. The resulting potential is an unbroken super- and parity-time ($\mathcal{PT}$)-symmetric shape-invariant potential with real energy eigenvalues, maintaining this property for all parameter values. In order to restore the probabilistic interpretation within a true quantum theory, a new inner product called the $\mathcal{CPT}$-inner product is defined in $\mathcal{PT}$-symmetric quantum mechanics, replacing the Dirac Hermitian inner product. In this work, we propose a new version of the inner product called the anti-$\mathcal{PT}$ ($\mathcal{APT}$)-inner product, $\langle A|B\rangle\equiv |A\rangle^{\mathcal{APT}}.|B\rangle$, which replaces the previous versions without any additional considerations. This $\mathcal{PT}$-supersymmetric quantum mechanics framework also allows for the unification of various areas of physics, including classical optics and quantum mechanics. To validate the theory, we present exact solutions for optical waveguides and the quantum tunneling probability, demonstrating excellent agreement with experimental data for the probability of crossing the potential barrier in the $\rm ^{3}H(d,n)^{4}He$ reaction.
\end{abstract}
\maketitle

%
\section{Introduction}
Space-time symmetries reflect the fundamental properties of any physical system and provide deep insights into the physics of the system. Since the discovery of the importance of combined space-time transformations, extensive theoretical and experimental research has been carried out on classical and quantum systems described by non-Hermitian Hamiltonians \cite{Ben98, Chen17, Feng17, Gana18, STARKOV2023169268}. The successes obtained have led to the development of a new branch of quantum mechanics (QM) called $\mathcal{PT}$-symmetric QM \cite{Ben07}. According to $\mathcal{PT}$-symmetric QM, a dynamically balanced open system has an unbroken $\mathcal{PT}$ symmetry and possesses real energy eigenvalues \cite{Ben18}. The applications of $\mathcal{PT}$-symmetric Hamiltonians have increased in various areas of physics, such as quantum field theory, condensed matter, quantum optics, and non-equilibrium statistical physics, since their introduction \cite{Bender18, Feng13, Hodaei14, Wen2021, Wu2019}. However, conventional quantum mechanics (CQM) is Hermitian, and thus fundamental issues related to probabilistic interpretation such as Hilbert space, positive definite norm, metric, unitary, and inner product become controversial when dealing with non-Hermitian Hamiltonians.

On the other hand, $\mathcal{APT}$ symmetry has also been the subject of theoretical and experimental researches in recent years \cite{Berg21, Choi18, Fang2021}. Being complex counterparts, these systems exhibit exclusively imaginary energy eigenvalues and adhere to $\left\{PT,H\right\}=0$. Consequently, an $\mathcal{APT}$-symmetric Hamiltonian, $H^{(\mathcal{APT})}$, can be mathematically obtained from a $\mathcal{PT}$-symmetric Hamiltonian, $H^{(\mathcal{PT})}$, by multiplying it by the imaginary number 'i', such that $H^{(\mathcal{APT})}=\pm iH^{(\mathcal{PT})}$ \cite{Peng16}. In spite of the theoretical elegance and empirical validations of $\mathcal{PT}$ and $\mathcal{APT}$ symmetries, these combined space-time transformations are challenging to intuitively understand. In order to better understand and study the relationship between these transformations, we employ them here for a solvable potential within QM.

The term "solvability" refers to a potential where the energy eigenvalues, eigenfunctions of the bound states, and the scattering matrix can be determined in closed analytic form. Supersymmetric quantum mechanics (SUSY-QM), as a formalism of QM, proposes a mechanism for generating and classifying solvable potentials \cite{Coo95}. Based on this framework, shape-invariant potentials (SIPs) form the main class of exactly solvable potentials. Moreover, within SUSY-QM, it has been demonstrated that the supersymmetric partner potentials (superpartners) $V_{1}(x)$ and $V_{2}(x)$ are isospectral, except for the lowest energy state of $V_{1}(x)$, by introducing the Hamiltonian hierarchy \cite{Gan17}. SUSY was originally introduced in the context of string and quantum field theories to unify the mathematical treatment of bosons and fermions \cite{Din07}. Although its experimental validation in particle physics is yet to be achieved, its application is rapidly expanding across many diverse areas of modern physics \cite{Kooh22, Kooh23, Hokm19, Ezaw23}. In this study, we introduce a new formalism to generate complex non-Hermitian $\mathcal{PT}$-symmetric superpartners, and then analyze them under $\mathcal{APT}$ transformations.

The remainder of the paper is organized as follows: In Sec. II, we provide a brief discussion on $\mathcal{PT}$ symmetry and SUSY-QM basics. Section III presents a new algebraic formalism to complexify the simplest solvable potential, namely the square-well potential, so that this formalism can be readily explored with other SIPs. The Hamiltonian hierarchy is used to obtain the energy spectrum and eigenfunctions in Sec. IV. Section V discusses $\mathcal{PT}$ and $\mathcal{APT}$ transformations by separating the odd and even terms of the potential and the superpotential. Sec. VI and Sec. VII examine two applications: one in classical optics, specifically optical waveguides, and the other in QM, specifically quantum tunneling probability. In Sec. VIII, we propose an $\mathcal{APT}$-inner product, an $\mathcal{APT}$-conjugation, as well as delve into the unitary time evolution. Finally, Sec. IX provides a conclusion.


 \section{$\mathcal{PT}$-Symmetry and SUSY-QM}
Since its inception, $\mathcal{PT}$-symmetric quantum mechanics has encountered challenges in some fundamental concepts, particularly those related to the probabilistic interpretation of wavefunctions. Addressing these challenges is crucial for $\mathcal{PT}$-symmetric QM to become a fully developed physical theory. One such challenge arises from the $\mathcal{PT}$-symmetric inner product, which can produce norms with indefinite signs and has been studied as pseudo-norms \cite{Levai2006}. To address this problem, a novel inner product called the $\mathcal{CPT}$-inner product was introduced, and its determination was made dynamically. This inner product allows a complex non-Hermitian Hamiltonian to possess an unbroken $\mathcal{PT}$ symmetry with a real and positive definite norm \cite{Bender2002}. The $\mathcal{C}$ operator is defined as a linear operator that commutes with both the Hamiltonian and $\mathcal{PT}$ operators, revealing a new symmetry of the Hamiltonian. However, calculating $\mathcal{C}$ in closed form even for solvable potentials is nontrivial, and explicit knowledge of $\mathcal{C}$ is only available in a few cases.

$\mathcal{PT}$-symmetric QM can be viewed as a generalization of CQM to the complex domain. However, understanding a physical system governed by a complex potential is more challenging intuitively compared to one governed by a real potential. Therefore, the study of exactly solvable systems is crucial for improving our understanding of the detailed structure of $\mathcal{PT}$-symmetric QM. Among all solvable models, SUSY-QM is recognized as a powerful method for recognizing, categorizing, obtaining, and generating solvable potentials. The techniques of SUSY-QM are based on the factorization method, which introduces ladder operators as (for simplicity, we set $\hbar = 1$ and $2m = 1$),
\begin{equation}
\left \{\begin{array}{ll} A_{n}=\frac{d}{dx}+W_{n}(k,x) \\
A^{\dag}_{n}=-\frac{d}{dx}+W_{n}(k,x). \end{array}
\right.
\end{equation}
Here, the subscript "$n$" indicates the nth member of the Hamiltonian hierarchy and $W_{n}(k,x)$ represents the superpotential corresponding to a function of main variable $x$ and a parameter $k$. The product of the operators generates partner Hamiltonians as,
\begin{equation}
\left \{\begin{array}{ll} H_{n}=A^{\mathcal{\dag}}_{n}A_{n}=-\frac{d^{2}}{dx^{2}}+V_{n}(k,x) \\
H_{n+1}=A_{n}A^{\mathcal{\dag}}_{n}=-\frac{d^{2}}{dx^{2}}+V_{n+1}(k,x), \end{array}
\right.
\end{equation}
so that $A_{n}$ and $A^{\mathcal{\dag}}_{n}$ are Dirac Hermitian conjugation (combined complex conjugation and matrix transposition) operators of each other. Based on equations (1) and (2), it seems that superpartners originated from the superpotential through the Riccati equation,
\begin{equation}
\left \{\begin{array}{ll} V_{n}(k,x)=W^{2}_{n}(k,x)- W^{\prime}_{n}(k,x)+E^{(n)}_{0}\\V_{n+1}(k,x)=W^{2}_{n}(k,x)+ W^{\prime}_{n}(k,x)+E^{(n)}_{0}, \end{array}
\right.
\end{equation}
where prime denotes $\partial/\partial x$ and $E^{(n)}_{0}$ is the ground-state energy of the potential $V_{n}$. The relations highlighted above emphasize the mutual influence between the space-time properties of the superpotential and the superpartners when considering quantum systems from this perspective. In $\mathcal{PT}$ symmetry, the real and imaginary components of the potential are even and odd, respectively. Given the discussions thus far, there must be a correlation between the even and odd properties of superpartners and superpotentials. This connection can be explored by separating the real and imaginary terms of the superpotential \cite{Levai2004}. Starting with a simple solvable example is often the most common approach to understanding a new subject and establishing valid rules. Accordingly, we also begin with the square-well problem.


 \section{$\mathcal{PT}$-symmetric Square-Well Superpartners}
Up until now, numerous variations of non-Hermitian potentials have been introduced, and their properties and practical applications have been discussed. Notably, by combining the concepts of SUSY and $\mathcal{PT}$ symmetry, complex non-Hermitian potentials have been discovered. Building upon these developments, we put forth a new algebraic formalism to introduce complexity to a well-known SIP, namely the square-well potential. This formalism allows for straightforward exploration of other SIPs as well.

The solutions of time-independent of Schr\"{o}dinger equation (SE) for ground-state of the simplest one-dimensional problem in CQM, i.e., particle in an infinite square-well potential with width $L$,
\begin{equation}
V_{1}(k,x)=\left \{\begin{array}{ll} -k^{2}~~~~-L/2< x < L/2 \\ \infty~~~~~~~ x\leq-L/2~\textrm{and}~x\geq L/2, \end{array}
\right.
\end{equation}
is known as trigonometric functions,
\begin{equation}
\left \{\begin{array}{ll} \psi_{0c}(k,x)=A\sin(kx) \\ \psi_{0t}(k,x)=B\cos(kx), \end{array}
\right.
\end{equation}
where $k$ is the wave number. In SUSY-QM, the nth-superpotential is defined as the logarithmic derivative of the nth ground state wave function \cite{Gan17},
\begin{equation}
W_{n}(k,x)=-\frac{d}{dx}\ln{\psi^{(n)}_{0}(k,x)}~;~~~~n=1,2,3,...
\end{equation}
Therefore, the superpotentials for the wave functions of the Eq. (5) are the cotangent (denoted by subscript $"c"$) and the tangent (denoted by subscript "t") functions, respectively. Now, to extend the superpotentials to the complex domain, we add an arbitrary imaginary functions, linearly,
\begin{equation}
\left \{\begin{array}{ll} W_{1c}(k,X)= -k\cot(X)+\textrm{i}W_{1ci}(X) \\
W_{1t}(k,X)=k\tan(X)+\textrm{i}W_{1ti}(X).
\end{array}
\right.
\end{equation}
where, $X=kx$ is the dimensionless length. For $n=1$ and due to unbroken SUSY, $E^{(1)}_{0}=0$, by putting the superpotentials $W_{1c}(k,X)$ and $W_{1t}(k,X)$ to Eq. (3), we get for cotangent and tangent functions, respectively,
\begin{equation}
\left \{\begin{array}{ll}  V_{1c}(k,X)=k(k- X^{\prime})\csc^{2}(X)-k^{2}-W^{2}_{1ci}(X)+\textrm{i}\left[-W^{\prime}_{1ci}(X)-2k\cot(X)W_{1ci}(X)\right]\\V_{2c}(k,X)=k(k+ X^{\prime})\csc^{2}(X)-k^{2}-W^{2}_{1ci}(X)+\textrm{i}\left[W^{\prime}_{1ci}(X)-2k\cot(X)W_{1ci}(X)\right], \end{array}
\right.
\end{equation}
and,
\begin{equation}
\left \{\begin{array}{ll}  V_{1t}(k,X)=k(k- X^{\prime})\sec^{2}(X)-k^{2}-W^{2}_{1ti}(X)+\textrm{i}\left[-W^{\prime}_{1ti}(X)+2k\tan(X)W_{1ti}(X)\right]\\V_{2t}(k,X)=k(k+ X^{\prime})\sec^{2}(X)-k^{2}-W^{2}_{1ti}(X)+\textrm{i}\left[W^{\prime}_{1ti}(X)+2k\tan(X^{\prime})W_{1ti}(X)\right], \end{array}
\right.
\end{equation}
If the $V_{1}$ and $V_{2}$ potentials are similar in shape and differ only in the parameters that appear in them, then they are said to shape invariant \cite{Gendenshtein1983}. The remainder that is defined as $R_{1}=V_{2}(k,X)-V_{1}(k+X^{\prime} ,X)$, equals with,
\begin{equation}
\left \{\begin{array}{ll} R_{1c}=X^{\prime}(X^{\prime}+2k)+\textrm{i}2\left[W^{\prime}_{1ci}(X)+X^{\prime}\cot(X)W_{1ci}(X)\right]\\
R_{1t}=X^{\prime}(X^{\prime}+2k)+\textrm{i}2\left[W^{\prime}_{1ti}(X)-X^{\prime}\tan(X)W_{1ti}(X)\right], \end{array}
 \right.
\end{equation}
The remainder, which gives the energy difference between the ground state energies of the partner Hamiltonians $H_{1}$ and $H_{2}$, i.e., $R_{1}=E^{(2)}_{0}-E^{(1)}_{0}$, is equal to the energy difference of the first two levels (the ground state and the first excited state) of the $H_{1}$, i.e., $R_{1}=E^{(1)}_{1}-E^{(1)}_{0}$. Consequently, the realness of $R_{1}$ guarantees that the both energy eigenvalues $E^{(1)}_{0}=0$ and $E^{(1)}_{0}=R_{1}$ are also real. If we continue this process with the Hamiltonian hierarchy method, we can be sure that the entire spectrum of the Hamiltonian $H_{1}$ is real. Moreover, superpartners are shape invariant only if the bracket terms of the remainders to be zero. As a result, the functions $W_{1ci}(X)$ and $W_{1ti}(X)$ are determined by this constraint as,
\begin{equation}
\left \{\begin{array}{ll} W_{1ci}(x)=q\csc(X)\\W_{1ti}(x)=q\sec(X), \end{array}
\right.
\end{equation}
where $q$ is an arbitrary constant. By setting the $X^{\prime}=k$, the superpotentials are gained as,
\begin{equation}
\left \{\begin{array}{ll} W_{1c}(k,X)=-k\cot(X)+\textrm{i} q\csc(X) \\ W_{1t}(k,X)=k\tan(X)+\textrm{i} q\sec(X), \end{array}
\right.
\end{equation}
The real and imaginary terms $W_{1cr}$ ($W_{1tr}$) and $W_{1ci}$ ( $W_{1ti}$) are plotted Fig. (1c), for typically value $q=1$. Proportionally, the complex non-Hermitian superpartners are,
\begin{equation}
\left \{\begin{array}{ll}  V_{1c}(k,X)=-q^{2}\csc^{2}(X)-k^{2}-\textrm{i}qk\cot(X)\csc(X)\\
V_{2c}(k,X)=(2k^{2}-q^{2})\csc^{2}(X)-k^{2}-\textrm{i}3qk\cot(X)\csc(X), \end{array}
\right.
\end{equation}
and,
\begin{equation}
\left \{\begin{array}{ll}  V_{1t}(k,X)=-q^{2}\sec^{2}(X)-k^{2}+\textrm{i}qk\tan(X)\sec(X)\\
V_{2t}(k,X)=(2k^{2}-q^{2})\sec^{2}(X)-k^{2}+\textrm{i}3qk\tan(X)\sec(X). \end{array}
\right.
\end{equation}
The potential $V_{1c}=V_{1t}=V_{1}=-1$ ($q=0$ and $k=1$) as well as real and imaginary parts $V_{1cr}$ ($V_{1tr}$) and $V_{1ci}$ ( $V_{1ti}$) are depicted in Fig. (1a). Moreover, the Fig. (1b) illustrates the potential $V_{2c}$ ($V_{2t}$) ($q=0$ and $k=1$)  as well as real and imaginary parts $V_{2cr}$ ($V_{2tr}$) and $V_{2ci}$ ( $V_{2ti}$), for typically values $q=1$ and $k=1$.
Finally, the wave functions are obtained by replacing superpotentials (12) in Eq. (6) as,
\begin{equation}
\left \{\begin{array}{ll} \psi^{(1)}_{0c}(k,X)=A|\sin(X)|\exp\left\{-f^{(1)}_{0c}\right\}~~~;~~~f^{(1)}_{0c}=-\textrm{i} \frac{q}{k}\ln\left[\csc(X)+\cot(X)\right] \\ \psi^{(1)}_{0t}(k,X)=B|\cos(X)|\exp\left\{-f^{(1)}_{0t}\right\}~~~;~~~f^{(1)}_{0t}=\textrm{i} \frac{q}{k}\ln\left[\sec(X)+\tan(X)\right], \end{array}
\right.
\end{equation}
The coefficients $A$ and $B$ are normalization constants. In Figure 2, the real and imaginary components of these wave functions are depicted.


 \section{Energy Eigenvalues and Eigenfunctions}
Now we consider the square-well in Eq. (4). The boundary conditions for the tangent function,
\begin{equation}
\psi(-\pi/2)=\psi(\pi/2)=0,
\end{equation}
imply that,
\begin{equation}
k_{n}=n+1
\end{equation}
where $n=0,1,2,...$ . For the cotangent function, the procedure is analogous, except for the modification that the well's width is shifted by $\pi/2$, and thus, $\psi(0)=\psi(\pi)=0$. As a result, the remainders (10) are,
\begin{equation}
R_{nt}=R_{nc}=(2n+1)k^{2}_{n},
\end{equation}
and according the unbroken SUSY, $E^{(1)}_{0}=0$, the energy spectrum is,
\begin{equation}
E^{(1)}_{n}=k^{2}_{n}-1=n(n+2)~;~ E^{(1)}_{n}=E^{(n+1)}_{0}
\end{equation}
The Hamiltonian hierarchy yields nth superpotentials, superpartners, and eigenfunctions, respectively as follow,
\begin{equation}
\left \{\begin{array}{ll} W_{nc}(k,x)= -nk_{n}\cot(k_{n}x)+\textrm{i}q\csc(k_{n}x) \\
W_{nt}(k,x)=nk_{n}\tan(k_{n}x)+\textrm{i}q\sec(k_{n}x),
\end{array}
\right.
\end{equation}
\begin{equation}
\left \{\begin{array}{ll}  V_{nc}(k,x)=\left[n(n-1)k^{2}_{n}-q^{2}\right]\csc^{2}(k_{n}x)-k^{2}_{n}-\textrm{i}(2n-1)qk_{n}\cot(k_{n}x)\csc(k_{n}x)\\
V_{nt}(k,x)=\left[n(n-1)k^{2}_{n}-q^{2}\right]\sec^{2}(k_{n}x)-k^{2}_{n}+\textrm{i}(2n-1)qk_{n}\tan(k_{n}x)\sec(k_{n}x). \end{array}
\right.
\end{equation}
and
\begin{equation}
\left \{\begin{array}{ll} \psi^{(n)}_{0c}(k,x)=A|\sin^{n}(k_{n}x)|\exp\left\{-f^{(n)}_{0c}\right\}~~~;~~~f^{(n)}_{0c}=-\textrm{i} \frac{q}{k_{n}}\ln\left[\csc(k_{n}x)+\cot(k_{n}x)\right]  \\ \psi^{(n)}_{0t}(k,x)=B|\cos^{n}(k_{n}x)|\exp\left\{-f^{(n)}_{0t}\right\}~~~;~~~f^{(n)}_{0t}=\textrm{i} \frac{q}{k_{n}}\ln\left[\sec(k_{n}x)+\tan(k_{n}x)\right], \end{array}
\right.
\end{equation}
If we set $q = 0$, these complex functions reduce to the their well-known real forms. Therefor, we can interpret that known real functions are, in fact, particular cases of complex versions, when the imaginary terms are zero.


 \section{$\mathcal{PT}$ and $\mathcal{APT}$ Transformations}
The $\mathcal{PT}$-symmetry requires that the potential under the following transformations be symmetric \cite{Ben98}:
\begin{equation}
\left \{\begin{array}{ll} x\rightarrow-x \\ t\rightarrow-t \\ i\rightarrow-i \end{array}
\right.
\end{equation}
This proves that the potentials are invariant under the simultaneous action of the space $\mathcal{P}$ and time $\mathcal{T}$ reflections, and have the property $V(x)=V^{*}(-x)$. In other words, this means the real $V_{r}(x)$ and imaginary $V_{i}(x)$ components of the potentials have to be even and odd functions of $x$, respectively \cite{ABHINAV2013110},
\begin{equation}
V^{\mathcal{PT}}(x)=V(x)=V^{even}_{r}(x)+\textrm{i}V^{odd}_{i}(x),
\end{equation}
Fig. (1a) shows that the potentials $V_{1tr}$ and $V_{1ti}$ and also Fig. (1b) illustrates that $V_{2tr}$ and $V_{2ti}$ satisfy these conditions (preserve $\mathcal{PT}$ symmetry). These complex $\mathcal{PT}$-symmetric superpartners are isospectrum with infinite square-well (except ground states). Therefore, only the tangent wave functions $\psi^{(n)}_{0t}$, corresponding to the bound states. If we shift the width of the well from the range $-\pi/2\leq L\leq \pi/2$ to the range $0\leq L\leq \pi$, we have to replace the tangent with the cotangent functions in order that invariance is preserved under the $\mathcal{PT}$ transformations.

Nevertheless, the superpotential $W_{1t}$ have anti-symmetric features, in the sense that the real $W_{1tr}$ and imaginary $W_{1ti}$ terms of the superpotential are odd and even functions, respectively (see Fig. (1c)). Hence, it can be inferred that $\mathcal{PT}$-symmetric superpartners originate from an underlying $\mathcal{APT}$-symmetric superpotential,
\begin{equation}
W^{\mathcal{APT}}_{1t}(x)=W_{1t}(x)=W^{odd}_{1tr}(x)+\textrm{i}W^{even}_{1ti}(x),
\end{equation}
where ${\mathcal{PT},W_{1t}}=0$. This is due to the change in the sign of the first spatial derivative, under $\mathcal{APT}$ transformations. Moreover, to preserve the potential algebra in SUSY-QM formalism, it is necessary the $\mathcal{APT}$-conjugation be similar to the Hermitian conjugation on this operator,
\begin{equation}
\left \{\begin{array}{ll} \left(\frac{d}{dx}\right)^{\mathcal{APT}}\rightarrow-\frac{d}{dx} \\ \left(\frac{d}{dx}\right)^{\dag}\rightarrow-\frac{d}{dx} \end{array}
\right.
\end{equation}
The same argument also can be applied to integration. This is because each derivative or integration operation can exchange the odd and even properties of a function. Now, by examining the Hermitian and $\mathcal{APT}$ conjugations on the ladder operators, we have,
\begin{equation}
\left \{\begin{array}{ll} A_{1t}=\frac{d}{dx}+W_{1t}(x)=\frac{d}{dx}+W_{1tr}(x)+\textrm{i}W_{1ti}(x) \\
A^{\mathcal{APT}}_{1t}=-\frac{d}{dx}+W^{\mathcal{APT}}_{1t}(x)=-\frac{d}{dx}+W_{1tr}(x)+\textrm{i}W_{1ti}(x) \\
A^{\dag}_{1t}=-\frac{d}{dx}+W^{\dag}_{1t}(x)=-\frac{d}{dx}+W_{1tr}(x)-\textrm{i}W_{1ti}(x) \end{array}
\right.
\end{equation}
In contrast to $\mathcal{APT}$ conjugation, we see that the Hermitian conjugation operator fails potential algebra (used in Sec. III) because it changes the sign of the superpotential imaginary term. Therefore, this mathematical operator is not longer applicable for a $\mathcal{PT}$-SUSY-QM.

So far, the $\mathcal{APT}$ transformations and symmetry are analyzed for a $\mathcal{PT}$-SUSY quantum system where both $\mathcal{PT}$ and SUSY symmetries remain unbroken. However, unlike $\mathcal{PT}$ transformations, the $\mathcal{APT}$ transformations are less intuitive. Therefore, we cannot exhibit their effects on individual quantities in a similar manner as shown in transformations (23). However, we can categorize  $\mathcal{PT}$ and  $\mathcal{APT}$ transformations of a typical function $g(x)$ as follows,
\begin{equation}
\left \{\begin{array}{ll}
\left[\textrm{i}g^{even}(x)\right]^{\mathcal{APT}}=-[\textrm{i}g^{even}(x)]^{\mathcal{PT}}=\textrm{i}g^{even}(x) \\
\left[\textrm{i}g^{odd}(x)\right]^{\mathcal{APT}}=-[\textrm{i}g^{odd}(x)]^{\mathcal{PT}}=-\textrm{i}g^{odd}(x) \end{array}
\right.
\end{equation}

Equivalently, it can also be formed by multiplying the potential by the imaginary unit '$\textrm{i}$' in a $\mathcal{PT}$-symmetric setting, $V^{\mathcal{APT}}(x)=\pm \textrm{i}V^{\mathcal{PT}}(x)$. This mathematical operation effectively swaps the properties of $\mathcal{PT}$ and $\mathcal{APT}$ symmetries within the potential. Furthermore, it can be deduced that an $\mathcal{APT}$-symmetric potential exclusively yields imaginary energy eigenvalues.


\section{Application 1: Optical Waveguides}
In the context of optics, SUSY has
been successfully employed for many applications in the study of
guided wave optics \cite{Vie21,Ward22}. In this section, we give a brief overview of the relationship between SUSY-QM and wave optics and then we get the refractive index (RI) from an optical potential (OP).
The SE for a particle propagation in a potential $V_{1}(x)$ and light propagation in an optical waveguide share a close analogy. In order to establish the analogy, consider the Helmholtz equation (HE) for the propagation of electromagnetic waves in the direction $z$.
The $E^{(1)}_{y}(x,z,t)$ component of the electric field is typically expressed as a superposition of all
the allowed modes, reading,
\begin{equation}
E^{(1)}_{y}(x,z,t)=\sum_{n}a^{(1)}_{n}e^{(1)}_{n}(x)\exp\left\{\textrm{i}(\beta^{(1)}_{n}z-\omega_{0}t)\right\}
\end{equation}
where $a^{(1)}_{n}$ is the amplitude, $e^{(1)}_{n}(x)$ the transverse electric (TE) field distribution, which is assumed to propagate
freely in the $z$ direction, $\beta^{(1)}_{n}$ the propagation constant of mode $n$, $\omega_{0}=ck_{0}$, $k_{0}=2\pi \lambda_{0}$ the wavenumber, $\lambda_{0}$ the wavelength, and $c$ is the speed of light in vacuum. Specifically, we have used the index "1" because of  Hamiltonian hierarchy in SUSY-QM allows us to generalize these equations directly to nth modes of mth waves. The HE describes the propagation of TE waves,
\begin{equation}
\left[-\frac{d^{2}}{dx^{2}}-k^{2}_{0}n_{1}^{2}(x)\right]e^{(1)}_{n}(x)=-\left(\beta^{(1)}_{n}\right)^{2}e^{(1)}_{n}(x)
\end{equation}
where $n_{1}(x)$ is the RI of the waveguide, $x$ is the confining transverse dimension.
The time-independent SE,
\begin{equation}
\left[-\frac{d^{2}}{dx^{2}}+V_{1}(x)\right]\psi^{(1)}_{n}(x)=E^{(1)}_{n}\psi^{(1)}_{n}(x),
\end{equation}
can be compared to Eq. (30) in which $e^{(1)}_{n}(x)$ plays the role of the wavefunction $\psi^{(1)}_{n}(x)$ of state $n$, $n_{1}(x)$ relates to the OP, $V_{1}(x)$,
\begin{equation}
V_{1}(x)=k^{2}_{0}\left[n^{2}_{0}-n_{1}^{2}(x)\right].
\end{equation}
as well as $\beta^{(1)}_{n}$ performs the role of the eigenenergy $E^{(1)}_{n}$,
\begin{equation}
E^{(1)}_{n}=k^{2}_{0}\left[n^{2}_{0}-n^{2}_{eff}\right].
\end{equation}
Here $n_{eff}=\beta^{(1)}/k_{0}$, and $n_{0}$ is any constant RI that can be
lucidly chosen to set $E^{(1)}_{0}=0$ to adapt with unbroken SUSY. According to Eqs. (32) and (33), the OP must also be a complex quantity with a real spectrum. Therefore, we have,
\begin{equation}
\left \{\begin{array}{ll} n_{1}(x)=\sqrt{n^{2}_{0}-k^{2}_{0}V_{1}(x)} \\
\beta^{(1)}_{n}=\sqrt{\beta^2_{0}-E^{(1)}_{n}} \end{array}
\right.
\end{equation}
where $\beta_{0}=n_{0}k_{0}$. In optics, the RI is complex, i.e., $n_{1}(x)=n_{1r}(x)+\textrm{i}n_{1i}(x)$, so its imaginary part indicates gain and loss. As a result, the real and imaginary terms of RI can be obtained by using polar coordinates as,
\begin{equation}
\left \{\begin{array}{ll} n_{1r}(x)=\rho_{1}(x)\cos\left[\theta_{1}(x)/2\right] \\
n_{1i}(x)=\rho_{1}(x)\sin\left[\theta_{1}(x)/2\right] \end{array}
\right.
\end{equation}
The final result is achieved by substituting this and $V_{1}(x)=V_{1r}(x)+\textrm{i}V_{1i}(x)$ into Eq. (34), which gives us the following relation,
\begin{equation}
\left \{\begin{array}{ll} \rho_{1}(x)=\left\{\left[n^{2}_{0}-k^{2}_{0}V_{1r}(x)\right]^{2}+k^{4}_{0}V^{2}_{1i}(x)\right\}^{1/4} \\
\theta_{1}(x)=\arctan\left[\frac{-k^{2}_{0}V_{1i}(x)}{n^{2}_{0}-k^{2}_{0}V_{1r}(x)}\right] \end{array}
\right.
\end{equation}
This formalism stipulates that $V_{1}(x)$ must be a complex potential with real eigenvalues. For example, if we put $V_{1t}(x)$ from Eq. (14), we have,
\begin{equation}
\left \{\begin{array}{ll} \rho_{1}(x)=\left\{\left[n^{2}_{0}+k^{2}_{0}(q^{2}\sec^{2}(kx)+k^{2})\right]^{2}+k^{4}_{0}[qk\tan(kx)\sec(kx)]^{2}\right\}^{1/4} \\
\theta_{1}(x)=\arctan\left[\frac{-k^{2}_{0}(qk\tan(kx)\sec(kx))}{n^{2}_{0}+k^{2}_{0}(q^{2}\sec^{2}(kx)+k^{2})}\right] \end{array}
\right.
\end{equation}
as well as $\beta^{(1)}_{n}=\sqrt{\beta^2_{0}-k^{2}_{n}-1}$ is resulted from inserting Eq. (19) into Eq. (34). One the benefits of this scheme, for instance, is the use of complex non-Hermitian $\mathcal{PT}$-supersymmetric RIs in order to selectively remove certain guided modes in optical waveguides \cite{Miri2013}.


\section{Application 2: Quantum Tunneling Probability (QTP)}
One of the most prominent examples of quantum tunneling is observed in nuclear fusion. Experimental observations indicate that light nuclei can overcome the Coulomb barrier even when their energy levels are below it. The nuclear optical theory, which employs a complex potential for the reaction, provides a satisfactory explanation for the fusion of light nuclei. Within this theory, the quantum tunneling probability (QTP) for the $\ell$-th partial wave can be expressed as \cite{Kooh16},
\begin{equation}
P_{\ell}(E)=\frac{4T\delta_{\ell i}}{T\delta^{2}_{\ell r}+(1+T\delta_{\ell i})^{2}}
\end{equation}
where the $T\delta_{r\ell}$ and $T\delta_{i\ell}$ are real and imaginary components of tangent of phase shift $\delta_{\ell}$, respectively,
\begin{equation}
\left \{\begin{array}{ll} \rm T\delta_{\ell r}=\frac{\rho \omega_{\ell r}
\left(F^{\prime}_{\ell}G_{\ell} +G^{\prime}_{\ell}F_{\ell}\right)
-G_{\ell} F_{\ell}\left(\omega_{\ell r}^2+\omega_{\ell
i}^2\right)-\rho^{2}F^{\prime}_{\ell}G^{\prime}_{\ell}}{G^{2}_{\ell}
 \left(\omega_{\ell r}^2+\omega_{\ell
i}^2\right)+\rho G^{\prime}_{\ell}\left(\rho G^{\prime}_{\ell}-2G_{\ell} \omega_{\ell r}\right)} \\
\rm T\delta_{\ell i}=\frac{-\rho \omega_{\ell i}}{G^{2}_{\ell}
 \left(\omega_{\ell r}^2+\omega_{\ell
i}^2\right)+\rho G^{\prime}_{\ell}\left(\rho G^{\prime}_{\ell}-2G_{\ell} \omega_{\ell r}\right)}
\end{array}
\right.
\end{equation}
where, $\rho=ka$, $k=\sqrt{2 \mu E/ \hbar^2}$ is the free particle wave number, $\mu$ is reduced mass,
$\eta=1 / ka_{C}$ is the Sommerfeld parameter,
$a_{C}=\hbar^2/\mu Z_{p}Z_{t}e^2$ is Coulomb unit length. Moreover,
$F_{\ell}(\eta,\rho)$ and $G_{\ell}(\eta,\rho)$ are regular and
irregular Coulomb wave functions and $Z_{p}e$ and $Z_{t}e$ are projectile and target charges,
respectively.
Considering the given argument, $\omega_{\ell r}(r)$ and $\omega_{\ell i}(r)$ in Eq. (39) represent the real and imaginary terms, respectively, of the superpotential. These terms are obtained by imposing boundary conditions. To ensure the continuity of the wave function and its first derivative, the logarithmic derivative of the wave functions at the boundary must match the conditions both inside and outside the boundary.
\begin{equation}
a\frac{\psi_{\textrm{int}}^\prime(r)}{\psi_{\textrm{int}}(r)}
 \bigg \vert_{r=a}=a\frac{\psi_{\textrm{ext}}^\prime(r)}{\psi_{\textrm{ext}}(r)} \bigg \vert_{r=a}
\end{equation}
In accordance with Eq. (6), the logarithmic derivative of a ground state wave function is the superpotential.
This implies that the superpotential is unique on all boundaries,
\begin{equation}
-aW_{\textrm{int}}(a)=\rho\frac{T
\delta_{\ell}G^{\prime}_{\ell}(\eta,\rho)+F^{\prime}_{\ell}(\eta,\rho)}{T
\delta_{\ell}G_{\ell}(\eta,\rho)+F_{\ell}(\eta,\rho)}
\end{equation}
where $T\delta_{\ell}\equiv\tan\delta_{\ell}=T\delta_{r\ell}+\textrm{i}T\delta_{i\ell}$ is tangent of phase shift. If we assume the nuclear potential as  $V_{1c}(r)$ from Eq. (13), within the range of $0\leq r\leq a$, the complex phase shift requires that the superpotential also to be complex,
\begin{equation}
-aW_{\textrm{int}}(a)=\omega_{\ell r}(a)+\textrm{i}\omega_{\ell i}(a)
\end{equation}
whence according to Eq. (12), we have,
\begin{equation}
\left \{\begin{array}{ll}  \omega_{\ell r}(a)=\rho_{N}\cot(\rho_{N})\\
\omega_{\ell i}(a)=-\rho_{q}\csc(\rho_{N})
\end{array}
\right.
\end{equation}
here $\rho_{N}=k_{N}a$, $\rho_{q}=qa$, $k_{N}=\sqrt{2 \mu (E-V_{0})/ \hbar^2}$ is the nuclear wave number and $V_{0}$ is depth of the potential. Substituting Eq. (43) in (39) and solving it leads to QTP (Eq. (38)) theoretically.

On the other hand, QTP relates to the reaction cross-section data via the following relation,
\begin{equation}
P_{\ell}(E)=\frac{k^2}{\pi g(I,s_{p},s_{t})(1+\delta_{pt})}\sigma_{re,\ell}(E)
\end{equation}
where $g(I,s_{p},s_{t})=(2I+1)/(2s_{p}+1)(2s_{t}+1)$ is statistical factor
dependent on spin of projectile, $s_{p}$, target, $s_{t}$, and
excited state in the compound nucleus $I$ and the Kronecker delta $\delta_{pt}$ is required to account for
the double counting of the identical particles ($p = t$).

The primary fuel for fusion reactions is deuterium-tritium due to its large fusion cross section $^{3}H(d,n)^{4}He$ at low energies ($E_{C.M}=40-100$ keV) and its high reactivity at achievable temperatures ($kT=1-30$ keV) \cite{Koohrokh16, Zyls22}. The low-energy cross section of $^{3}H+d$ is predominantly governed by an s-wave ($\ell=0$) resonance with a spin-parity of $I^{\pi}=3/2^{+}$ \cite{Kooh16}. The parameters $V_{0}=-36.215$ MeV, $q=0.01929~(\textrm{fm}^{-1})$, and $a=5.4497$ fm are determined by fitting Eq. (38) to the experimental data of $P_{0}(E)$, obtained from the fusion cross section using Eq. (44). The comparison between theoretical and experimental results for the QTP is shown in Fig. 3 \cite{Jar84, Bro87, Kob66, Arn53, Con52}. In this approach, the potential describing the compound nucleus has real energy eigenvalues, which makes the method superior to nuclear optical models. Moreover, the excellent agreement between the theoretical results and experimental data further validates the theory.


 \section{Probabilistic Interpretation of PT-SUSY-QM: $\mathcal{APT}$-Inner Product, $\mathcal{APT}$-Conjugation and Unitarity}
In line with the probabilistic interpretation, quantum systems possess an inherent probabilistic nature. Consequently, any valid physical quantum theory should accurately embody this probabilistic interpretation, wherein probabilities are linked to the results of measurements in the quantum realm. This challenge is something that $\mathcal{PT}$-symmetric QM has been grappling with ever since its inception. This implies that $\mathcal{PT}$-symmetric QM should possess a Hilbert space of state vectors, characterized by an inner product that has a positively definite norm (definite metrics). This arises from the fact that the norm of a state is understood as a probability, and probabilities inherently represent positive values. Additionally, the theory's Hamiltonian should induce unitary time evolution, as the preservation of probability over time necessitates this for a stable system. Here, a novel inner product is introduced, which serves as the foundation for restoring the accurate probabilistic understanding within the framework of $\mathcal{PT}$-SUSY-QM, for unbroken $\mathcal{PT}$ symmetry as well as unbroken SUSY systems.

\subsection{$\mathcal{APT}$-Inner Product and $\mathcal{APT}$-Conjugation}
In SUSY-QM, a ground-state wave function can be written in a general form \cite{Gan17} (Eq. 22),
\begin{equation}
\psi^{(n)}_{0}(x)=N\exp\{-f_{n}(x)\},
\end{equation}
where function $f_{n}(x)$ is obtained by integrating superpotential,
\begin{equation}
f_{n}(x)=\int W_{n}(x)dx,
\end{equation}
According to Sec. V arguments, in a $\mathcal{PT}$-SUSY system, $f_{n}(x)$ and $W_{n}(x)$ should be $\mathcal{PT}$- and $\mathcal{APT}$-symmetric, respectively,
\begin{equation}
\left \{\begin{array}{ll} [f_{n}(x)]^{\mathcal{PT}}=f_{n}(x)=f^{even}_{nr}(x)+\textrm{i}f^{odd}_{ni}(x) \\
\left[W_{n}(x)\right]^{\mathcal{APT}}=W_{n}(x)=W^{odd}_{nr}(x)+\textrm{i}W^{even}_{ni}(x) \end{array}
\right.
\end{equation}
With respect to SUSY-QM formalism, we know that supersymmetric eigenfunctions consist of a complete set and are orthonormal. Obviously, the $\mathcal{PT}$-SUSY eigenfunctions (22) posse also these properties. However, the coordinate-space inner product needs to redefine in such a way that the Hermitian inner product is replaced by the $\mathcal{APT}$-inner product as,
\begin{equation}
(\phi,\psi)=\int\left[\phi(x)\right]^{\mathcal{APT}}\psi(x)dx,
\end{equation}
where $[\mathcal{PT},\phi(x)]=[\mathcal{PT},\psi(x)]=0$. As a consequence, orthonormality reads,
\begin{equation}
(\psi^{(m)}_{0},\psi^{(n)}_{0})=\int[\psi^{(m)}_{0}(x)]^{\mathcal{APT}}\psi^{(n)}_{0}(x)dx=\delta_{m,n},
\end{equation}
and thus, the normalization constant is obtained by,
\begin{equation}
N=\frac{1}{\sqrt{\int \exp \{-2f^{even}_{nr}(x)\} dx}},
\end{equation}

From Eq. (27) we conclude that the first-order differential operators, named the ladder operators, should be redefined by $\mathcal{APT}$-conjugation as,
\begin{equation}
\left \{\begin{array}{ll} A=\frac{d}{dx}+W(x)=\frac{d}{dx}+W_{r}(x)+\textrm{i}W_{i}(x) \\
A^{\mathcal{APT}}=-\frac{d}{dx}+W(x)=-\frac{d}{dx}+W_{r}(x)+\textrm{i}W_{i}(x). \end{array}
\right.
\end{equation}
Accordingly, Hamiltonian superpartners,
\begin{equation}
\left \{\begin{array}{ll} H_{1}=A^{\mathcal{APT}}A \\
H_{2}=AA^{\mathcal{APT}}, \end{array}
\right.
\end{equation}
Therefore, the mathematical structure of $\mathcal{PT}$-SUSY-QM is as same as SUSY-QM, with difference that $\mathcal{APT}$-conjugation exchange with Hermitian conjugation.

\subsection{Unitarity}
We demonstrate that the $\mathcal{APT}$ inner product fulfills the necessary conditions for the quantum theory defined by a $\mathcal{PT}$-symmetric Hamiltonian to possess a Hilbert space with a positive norm and to operate as a unitary theory. Solving the time part of the SE leads to $\psi(t)\propto \exp\{-\textrm{i}Ht\}$. According to Sec. V, we have,
\begin{equation}
 \psi(t)=C\exp\left\{\left(-\textrm{i}p^{2}-\textrm{i}V^{even}_{r}(x)+V^{odd}_{i}(x)\right)t\right\},
\end{equation}
where $C$ is a real constant. This wave function is unequivocally $\mathcal{PT}$-symmetric, as it remains invariant under (23) and (28) transformations, meaning that $\psi^{\mathcal{PT}}(t)=\psi(t)$. During an $\mathcal{APT}$ transformation, we find that,
\begin{equation}
\psi^{\mathcal{APT}}(t)=C\exp\left\{\left(\textrm{i}p^{2}+\textrm{i}V^{even}_{r}(x)-V^{odd}_{i}(x)\right)t\right\},
\end{equation}
By utilizing the $\mathcal{APT}$ inner product, we find that the probability, denoted as $P=\psi^{\mathcal{APT}}\psi^{\mathcal{PT}}$, remains constant. When the probability remains unchanging over time, this reflects a unitary time evolution. Such a state characterizes an unbroken $\mathcal{PT}$-symmetric phase. Indeed, this system bears similarity to an isolated system typically described by a Hermitian Hamiltonian. However, it distinguishes itself as an open system in a state of dynamic equilibrium, where the rates of loss and gain are in perfect balance.

Now consider a wave function that is neither $\mathcal{PT}$-symmetric nor $\mathcal{APT}$-symmetric, e.g. Eq. (13), it follows $\psi(t) \propto \exp\left\{\left(\textrm{i}V^{even}_{r}(x)-V^{even}_{i}(x)\right)t\right\}$. The probability $P=\psi^{\mathcal{APT}}\psi$ gives,
\begin{equation}
P\propto \exp\left\{-2V^{even}_{i}(x)t\right\},
\end{equation}
In this scenario, it's important to note that the probability is not held constant, and thus, experiences exponential growth over time for negative values of $V^{even}_{i}(x)$ (gain), and undergoes exponential decay over time for positive values of $V^{even}_{i}(x)$ (loss). This scenario corresponds to an open system in a broken $\mathcal{PT}$-symmetric phase, characterized by an imbalance between gain and loss, leading to a non-equilibrium dynamic state.


 \section{Conclusion}
In fact, a real $\mathcal{PT}$-symmetric system refers to a system that maintains dynamic equilibrium (balanced gain and loss) while interacting with its external environment. At this point, the system is said to be in an unbroken $\mathcal{PT}$-symmetric phase, resulting in real energy eigenvalues. Conversely, a broken $\mathcal{PT}$-symmetric phase occurs when this dynamic equilibrium cannot be sustained, leading to complex eigenvalues. Hence, a $\mathcal{PT}$ transition can occur when an unbroken $\mathcal{PT}$ symmetry (equilibrium) transforms into a broken $\mathcal{PT}$ symmetry (non-equilibrium). The $\mathcal{PT}$ phase transition can happen suddenly or gradually, depending on the type of potentials and their parameter values. However, certain $\mathcal{PT}$-symmetric potentials do not exhibit $\mathcal{PT}$ transitions, and their energy eigenvalues remain real across the entire parameter range. In Section III, we derive potentials that preserve $\mathcal{PT}$ symmetry as well as SUSY for all parameter values within the real range. It should be noted that this is an example of $\mathcal{PT}$-SUSY SIPs, and this property can be sought for all other members of the SIPs family.

Now, let's consider two different approaches: CQM and the version of $\mathcal{PT}$-SUSY-QM presented in this article. These approaches aim to find a common solution and compare them. Typically, when solving the SE, the wave function is separated into two terms: regular and irregular. In CQM, the coefficients of these solutions are determined by boundary conditions. In many problems involving a singular point (often the origin), the irregular solution is discarded due to divergence, leaving only the regular term as the acceptable solution. In contrast, $\mathcal{PT}$-SUSY-QM automatically selects the regular term based on the $\mathcal{PT}$ symmetry criteria. This raises the question of whether $\mathcal{PT}$ symmetry arguments could also be beneficial in choosing the less singular solution within a singular potential \cite{Frank1971}.

Sections VI and VII present two examples from distinct fields of physics: one concerns an optical waveguide in classical mechanics, and the other pertains to QTP in QM. These examples illustrate the unifying role of the $\mathcal{PT}$-symmetric optical potential, which bridges the gap between these diverse fields. In the optical waveguide scenario, the RI emerges as crucial, being directly connected to the optical potential and consequently exhibiting $\mathcal{PT}$ symmetry (Eq. 37). On the other hand, in the context of QTP, the phase shift assumes significance, tied to the superpotential and displaying $\mathcal{APT}$ symmetry (Eq. 39).

The definition of the $\mathcal{CPT}$-inner product requires that the operator $\mathcal{C}$ commutes simultaneously with the Hamiltonian and $\mathcal{PT}$ operators. Consequently, the $\mathcal{C}$ operator implies an unforeseen symmetry of the Hamiltonian. For example, $\mathcal{C}$ is the complex extension of the
intrinsic parity operator in quantum field theory transforms under the Lorentz group as a scalar \cite{bender2005c}. However, what does such symmetry mean in non-relativistic quantum mechanics? Has it been observed in any experiments thus far? While $\mathcal{APT}$ symmetry is supported by some experimental evidence, and there is no need to calculate a new operator \cite{Bergman2021, Zhao2020, Fang2021, hu2023anti}.
\newpage
\begin{figure}
  \includegraphics[width=0.95\textwidth]{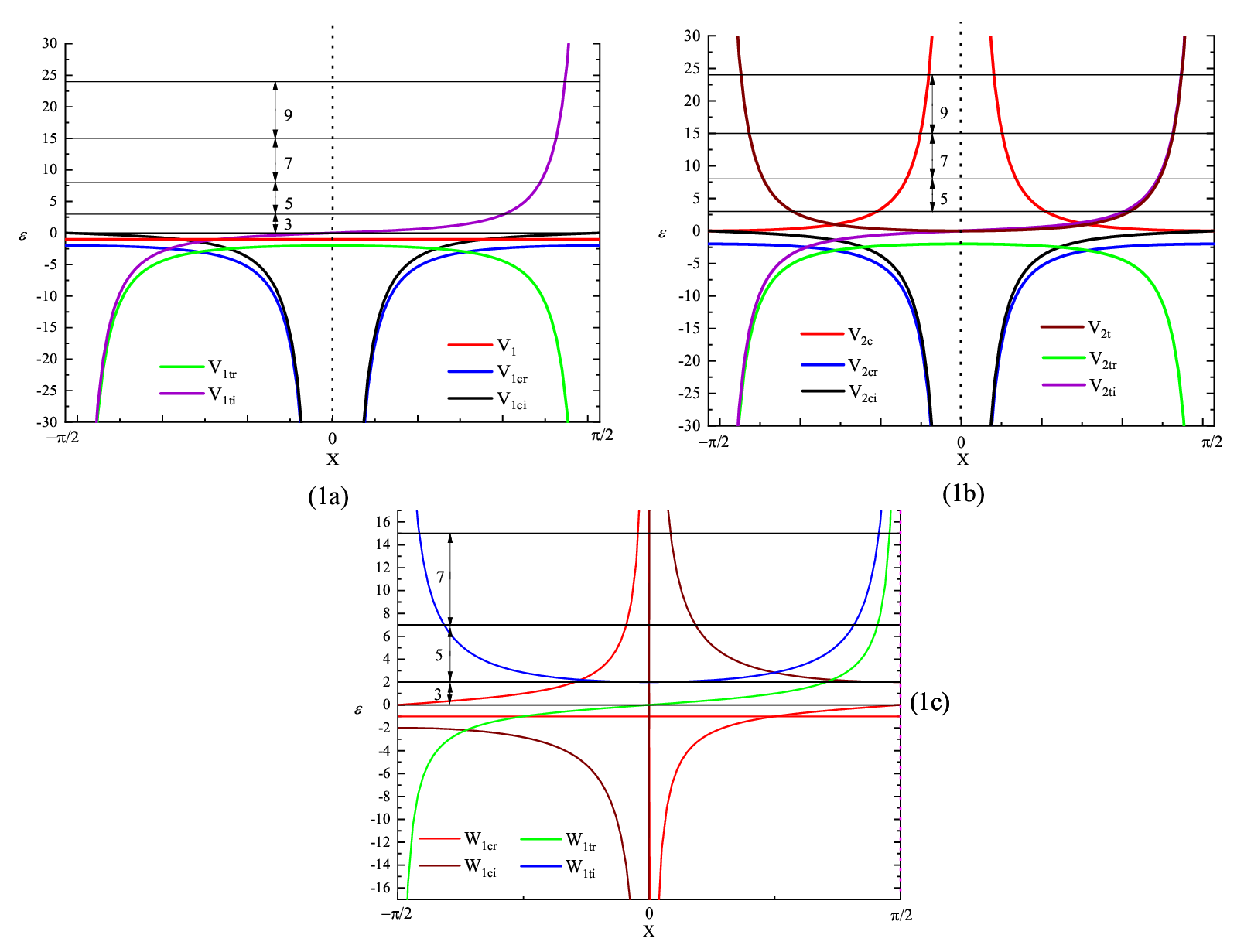}
\caption{1a: $V_{1}=-1$ and the real and imaginary terms are denoted as $V_{1cr}$ and $V_{1ci}$, and $V_{1tr}$ and $V_{1ti}$, respectively. 1b: $V_{2}$ is denoted as $V_{2c}$ and $V_{2t}$, and the real and imaginary terms are represented by $V_{2cr}$ and $V_{2ci}$, and $V_{2tr}$ and $V_{2ti}$, respectively. 1c: The real and imaginary terms are $W_{1cr}$ and $W_{1ci}$, and $W_{1tr}$ and $W_{1ti}$, respectively, for typically chosen values of $q=1$ and $k=1$. In the corresponding figures, the horizontal axis represents the dimensionless length $X=kx$, while the vertical axis represents the dimensionless energy $\epsilon=E/k^{2}$. } \label{fig:1}
\end{figure}
\clearpage


\begin{figure}
  \includegraphics[width=0.95\textwidth]{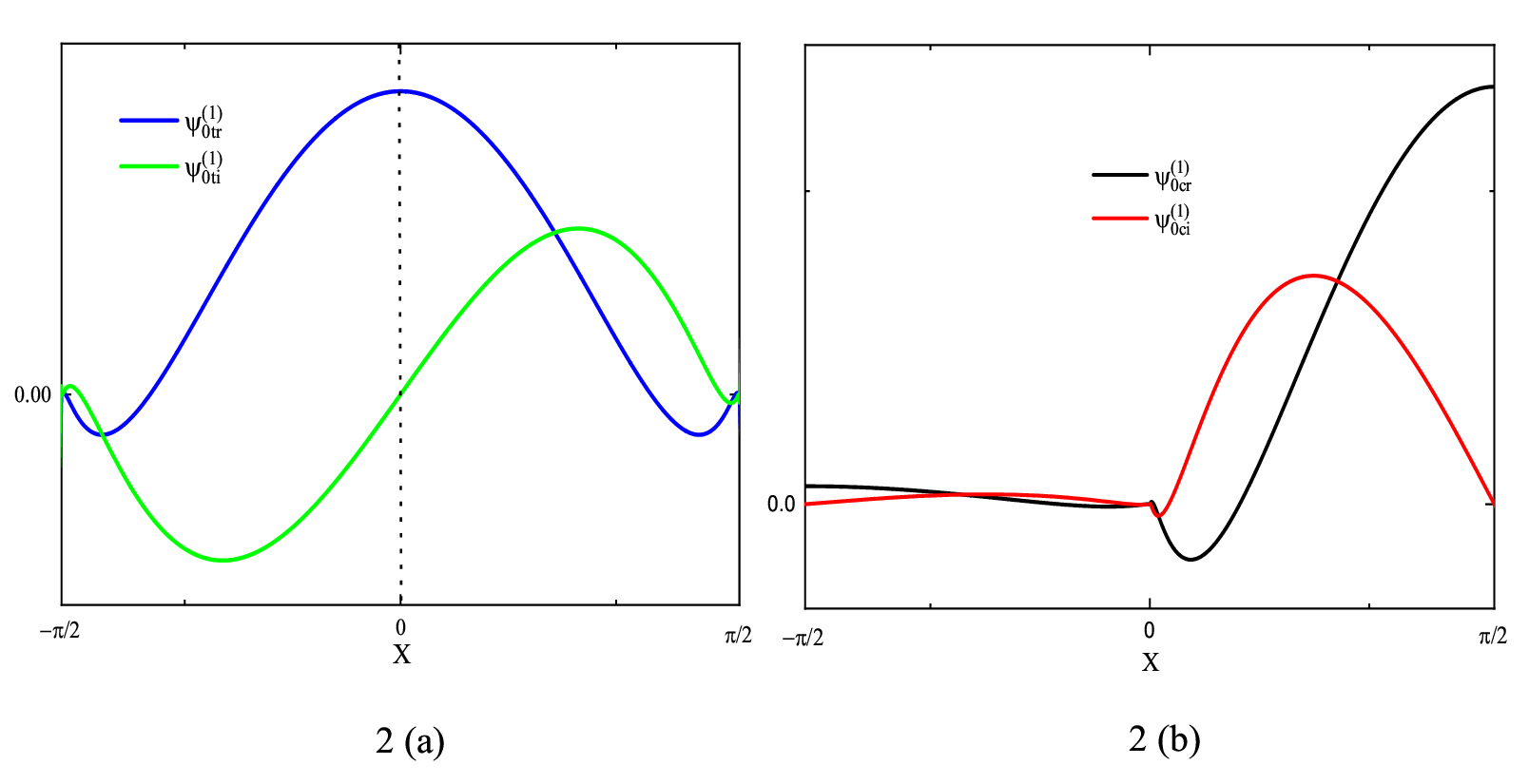}
\caption{(a) The real and imaginary terms of the tangent wave function and (b) The real and imaginary terms of the cotangent wave function (Eq. 15).}
\label{fig:2}
\end{figure}
\clearpage


\begin{figure}
  \includegraphics[width=0.95\textwidth]{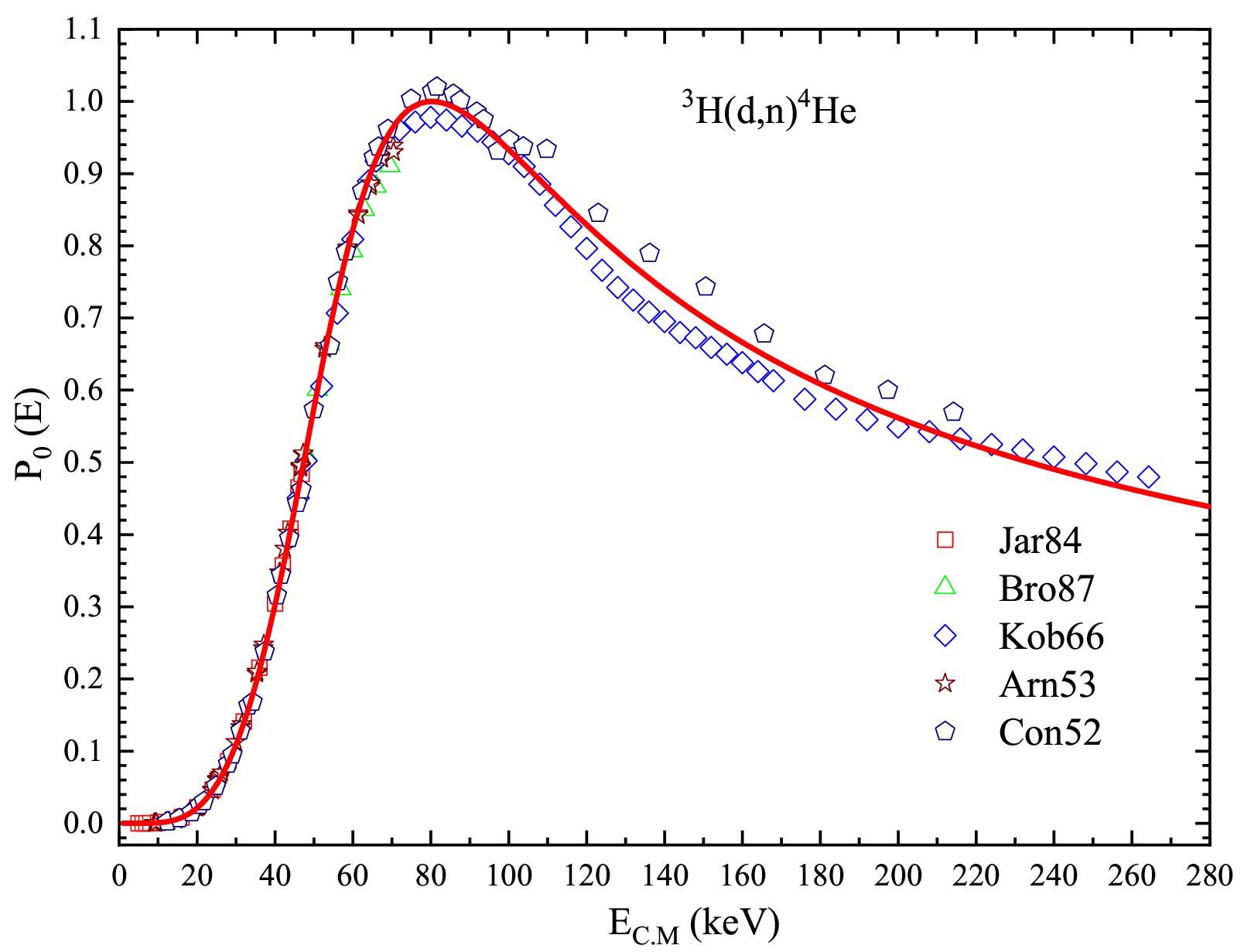}
\caption{The quantum tunneling probability (QTP) of the $\rm ^{3}H(d,n)^{4}He$ reaction is depicted in the figure. The solid line represents the theoretical result, while the scattered data points are obtained from experimental measurements.}
\label{fig:3}
\end{figure}
\clearpage


\end{document}